\documentclass[showpacs,twocolumn,amssymb]{revtex4}
\usepackage{graphicx}

\begin{document}

\title{Phase Diagram of Cold Polarized Fermi Gas in Two Dimensions}
\author{Lianyi He and Pengfei Zhuang}

\affiliation{Physics Department, Tsinghua University, Beijing
100084, China}

\begin{abstract}
The superfluid phase diagrams of a two-dimensional cold polarized
Fermi gas in the BCS-BEC crossover are systematically and
analytically investigated. In the BCS-Leggett mean field theory,
the transition from unpolarized superfluid phase to normal phase
is always of first order. For a homogeneous system, the two
critical Zeeman fields and the critical population imbalance are
analytically determined in the whole coupling parameter region,
and the superfluid-normal mixed phase is shown to be the ground
state between the two critical fields. The density profile in the
presence of a harmonic trap calculated in the local density
approximation exhibits a shell structure, a superfluid core at the
center and a normal shell outside. For weak interaction, the
normal shell contains a partially polarized cloud with constant
density difference surrounded by a fully polarized state. For
strong interaction, the normal shell is totally in fully polarized
state with a density profile depending only on the global
population imbalance. The di-fermion bound states can survive in
the whole highly imbalanced normal phase.
\end{abstract}

\pacs{03.75.Ss, 05.30.Fk, 74.20.Fg, 34.90.+q} \maketitle

\section {Introduction}
The effect of Zeeman energy splitting $h$ induced by a strong
magnetic field between spin up and down electrons on
Bardeen-Cooper-Shriffer(BCS) superconductivity, which has been
investigated many years ago~\cite{CC,Sarma,FFLO}, promoted new
interest in recent years due to the progress in the experiments of
ultracold Fermi gases\cite{exp1,exp2,exp3,exp4,exp5,exp6}. The
well-known result for weak-coupling s-wave superconductivity is
that, at a critical Zeeman field or the so-called
Chandrasekhar-Clogston(CC) limit $h_c=0.707\Delta_0$ where
$\Delta_0$ is the zero temperature gap, the Cooper pairs are
destroyed and a first order quantum phase transition from the
gapped BCS state to the normal state occurs\cite{CC}. Further
studies showed that the inhomogeneous
Fulde-Ferrell-Larkin-Ovchinnikov(FFLO) state\cite{FFLO} where
Cooper pairs have nonzero momentum can survive above the CC limit
up to $h_{\text{FFLO}}=0.754\Delta_0$. However, since the
thermodynamic critical field is much smaller than the CC limit due
to strong orbit effect\cite{CC}, it is hard to observe the CC
limit and the FFLO state in ordinary superconductors.

Recent experiments on ultracold Fermi gas trapped in an external
harmonic potential, serve as an alternative way to study the pure
Zeeman effect on Fermi
superfluidity~\cite{exp1,exp2,exp3,exp4,exp5,exp6}. The atom
numbers of the two lowest hyperfine states of $^6$Li atom,
$N_\uparrow$ and $N_\downarrow$, are adjusted to create a
population imbalance or polarization
$P=(N_{\uparrow}-N_{\downarrow})/(N_{\uparrow}+N_{\downarrow})$,
which simulates the Zeeman field $h$ in a superconductor. The
s-wave attraction between the two hyperfine states is tuned around
the Feshbach resonance to realize a strongly rather than weakly
interacting Fermi gas. In three-dimensional case, the density
profiles observed in experiments exhibit an unpolarized superfluid
core in the center of the trap and a polarized normal gas shell
outside\cite{exp1,exp2,exp3,exp4}, which justifies that the ground
state around unitary is a phase separation state(although there
may be some dispute on the shell structure), predicted by early
theoretical works\cite{BCR,Cohen,Carlson,Sheehy,MFwork}.

One of the theoretical interests in the study of polarized Fermi
superfluidity is to determine its phase structure in the whole
interaction strength region, namely in the BCS-BEC (Bose-Einstein
Condensation) crossover~\cite{Sheehy,MFwork}. The complete mean
field phase diagram in coupling-imbalance plane as well as the
critical Zeeman field $h_c$ and critical polarization $P_c$ are
theoretically predicted in three dimensional case
(3D)~\cite{Sheehy}. Since the s-wave mean field equations can not
be solved analytically in the whole coupling region, it is hard to
determine a precise phase diagram in 3D even in mean field
approximation. Recently, the quantitatively correct phase diagram
in 3D has been obtained in quantum Monte Carlo
calculations\cite{MC1,MC2}. However, the theoretical prediction of
the phase diagram for a homogeneous system can not be directly
examined in ultracold Fermi gas experiments, due to the effect of
the external harmonic trap. To have a comparison with the
experimental data, one should investigate the phase diagram and
the density profile in the presence of an external trap potential,
using the same equation of state. In the case of 3D, the density
profile can only be treated numerically\cite{trap}.

While we have well understood the 3D phase diagram, the phase
structure of polarized Fermi gases in low dimensions promoted
recently experimental and theoretical interests. In one
dimensional case the phase diagram is determined via exact
solvable models~\cite{1D}. In the two-dimensional case, while
exact solvable models are lacked, the s-wave mean field equations
can be solved analytically in the whole coupling parameter
region~\cite{2Da,2Db,2Dc,2Dreview}, and the Fermi surface topology
and stability condition are not trivial and different from those
in 3D~\cite{guba}. In this paper, we will determine the phase
diagrams of a polarized Fermi gas in two dimensions, for both
homogeneous and trapped systems, and calculate the density profile
of a trapped imbalance Fermi gas. Our results are totally
analytical in the whole coupling parameter region, including the
phase diagrams and the density profile. In the final part of this
paper, we also discuss the existence of di-fermion bound states in
the polarized normal phase.

\section {BCS-BEC Crossover in Two Dimensions}
The BCS-BEC crossover problem in two dimensions has been widely
discussed in the literatures\cite{2Da,2Db,2Dc,2Dreview}. In this
paper, we employ an effective 2D Hamiltonian where the
renormalized atom-atom interaction can be characterized by an
effective binding energy\cite{2Da}. For a wide Feshbach resonance,
the effective grand canonical Hamiltonian can be written as
\begin{eqnarray}
H&=&\sum_{\sigma=\uparrow,\downarrow}\int d^2{\bf
r}\psi_{\sigma}^\dagger({\bf
r})\left(-\frac{\hbar^2}{2M}\nabla^2-\mu-\sigma_zh\right)\psi_{\sigma}^{\phantom{\dag}}({\bf r})\nonumber\\
&-&U\int d^2{\bf r}\psi_{\uparrow}^\dagger({\bf
r})\psi_{\downarrow}^\dagger({\bf
r})\psi_{\downarrow}^{\phantom{\dag}}({\bf
r})\psi_{\uparrow}^{\phantom{\dag}}({\bf r}),
\label{Hamiltonian.2D}
\end{eqnarray}
where $M$ is the fermion mass, $\mu$ is the chemical potential,
$U>0$ is the contact attractive interaction, and $\sigma_z=\pm1$
correspond to $\sigma=\uparrow,\downarrow$. We choose the unit
$\hbar=1$ through the paper. The Zeeman field $h$ can be created
by either an external field\cite{CC,Sarma,FFLO} or a population
imbalance\cite{exp1,exp2}. In the former case, the total particle
number $N$ is conserved, but the particles $N_\uparrow$ and
$N_\downarrow$ in the states $\uparrow$ and $\downarrow$ can
transfer to each other\cite{liu}, i.e., the chemical potentials
for the two components are always the same, but the external field
$h$ induces an effective chemical potential difference. In the
latter case, $N_\uparrow$ and $N_\downarrow$ are both conserved,
and the two chemical potentials can be expressed as
$\mu_\uparrow=\mu+h$ and $\mu_\downarrow=\mu-h$.

At finite temperature in 2D, the long range order is absent and no
phase transition can happen. At zero temperature, however, there
do exist long range order~\cite{2Dreview} and one can safely
consider phase transitions among different states. In this paper,
we will study the phase diagrams at zero temperature in the
BCS-Leggett mean field approximation which is accepted to
adequately describe the BCS-BEC crossover at $T=0$~\cite{BCSBEC}.

In the balanced case with $h=0$, the thermodynamic potential
density of a uniform Fermi gas can be evaluated as\cite{2Dreview}
\begin{eqnarray}
\Omega(\mu;\Delta)=\frac{\Delta^2}{U}+\int\frac{d^2{\bf
k}}{(2\pi)^2}\left(\xi_{\bf k}-E_{\bf k}\right)
\end{eqnarray}
with the definition of particle energies $E_{\bf k}=\sqrt{\xi_{\bf
k}^2+\Delta^2}$ and $\xi_{\bf k}={\bf k}^2/(2M)-\mu$ and the
superfluid order parameter
$\Delta=-U\langle\psi_{\downarrow}^{\phantom{\dag}}\psi_{\uparrow}^{\phantom{\dag}}\rangle$.
In the dilute limit, the UV divergence in the expression of
$\Omega$ can be eliminated via introducing the two body scattering
length.

While the two-body bound state in 3D forms only at sufficiently
strong attraction where the s-wave scattering length diverges and
changes sign, the bound state in 2D can form at any arbitrarily
small attraction\cite{QM}. For an inter-atomic potential described
by a 2D circularly symmetric well of radius $R_0$ and depth $V_0$,
the bound-state energy $\epsilon_{\text B}$ is given by
$\epsilon_{\text B} \simeq 1/(2MR_0^2) \exp{[-2/(MV_0R_0^2)]}$
with $V_0 R_0^2 \to 0$. As a consequence, the solution of the
BCS-BEC problem in 2D is much simpler than that in the case of 3D
in terms of special functions\cite{2Db}. It is shown that the
existence of the two-body bound state in vacuum is a necessary
(and sufficient) condition for the Cooper instability\cite{2Da}.
To regulate the UV divergence in $\Omega$, we introduce a high
energy cutoff $\Lambda={\bf k}_\Lambda^2/(2M)$ in the integral.
The momentum cutoff ${\text k}_\Lambda$ corresponds to the inverse
of the range $r_0$ of the interaction potential. Due to the energy
independence of the density of states in 2D, after performing the
integration over ${\bf k}$ one obtains
\begin{eqnarray}
\Omega&=&\frac{\Delta^2}{U}-\frac{M\Delta^2}{4 \pi} \bigg[ \ln
\frac{\Lambda-\mu + \sqrt{(\Lambda-\mu)^{2}+\Delta^{2}}}
{\sqrt{\mu^{2}+\Delta^{2}}-\mu}\nonumber\\
&+&\frac{\Lambda-\mu}{\Lambda- \mu + \sqrt{(\Lambda- \mu)^{2} +
\Delta^{2}}} + \frac{\mu}{\sqrt{\mu^{2} + \Delta^{2}} - \mu}
\bigg].
\end{eqnarray}

In this paper we consider a dilute Fermi gas with effective
interaction range $r_0\rightarrow 0$. Taking large enough cutoff
$\Lambda$ and small enough attraction $U$, we can introduce a 2D
two-body binding energy\cite{2Dreview} to replace the cutoff in
this limit,
\begin{equation}
\epsilon_{\text B} = 2\Lambda \exp \left(-\frac{4 \pi}{M U}
\right),
\end{equation}
which does not include any many-particle effect. With the binding
energy, the cutoff dependence can be eliminated in the dilute
limit with $\Lambda \to \infty$ and $U \to 0$ but finite
$\epsilon_{\text B}$. We obtain in this limit
\begin{equation}
\Omega=\frac{M\Delta^2}{4\pi} \left( \ln \frac{\sqrt{\mu^{2} +
\Delta^{2}} - \mu}{\epsilon_{\text B}} - \frac{\mu}{\sqrt{\mu^{2}
+ \Delta^{2}} - \mu} - \frac{1}{2} \right) .
\end{equation}
The above procedure is equivalent to directly substituting the
coupling constant $U$ by the 2D bound state equation
\begin{equation}
\frac{1}{U}=\int\frac{d^2{\bf k}}{(2\pi)^2}\frac{1}{{\bf
k}^2/M+\epsilon_{\text{B}}}.
\end{equation}

The BCS-BEC crossover phenomenon in 2D can be observed by solving
the coupled gap and number equations, namely
$\partial\Omega/\partial\Delta=0$ and
$n=-\partial\Omega/\partial\mu$. Defining the Fermi energy
$\epsilon_{\text F}=\pi n/M$ in 2D, the gap and number equation
can be analytically expressed as
\begin{equation}
\sqrt{\mu^{2} + \Delta^{2}} - \mu = \epsilon_{\text B}, \qquad
\sqrt{\mu^{2} + \Delta^{2}} + \mu = 2\epsilon_{\text F},
\end{equation}
respectively. Their solution takes a very simple form\cite{2Da}
\begin{equation}
\Delta_0 = \sqrt{2 \epsilon_{\text B} \epsilon_{\text F}} \,,
\qquad \mu_0 = \epsilon_{\text F}- \frac{\epsilon_{\text B}}{2} ,
\end{equation}
or rewrite it in terms of a dimensionless quantity
$\eta=\epsilon_{\text B}/\epsilon_{\text F}$,
\begin{equation}
\frac{\Delta_0}{\epsilon_{\text F}}=\sqrt{2\eta} \,, \qquad
\frac{\mu_0}{\epsilon_{\text F}}=1-\frac{\eta}{2} .
\end{equation}
One sees very clear that the chemical potential decreases with
increasing coupling or decreasing density, which indicates a
BCS-BEC crossover. The Chemical potential changes sign at
$\epsilon_{\text B}=2\epsilon_{\text F}$. To understand the
physical significance of these simple results, we consider two
limits. For very weak attraction (or high density), the
two-particle binding energy is extremely small, i.e.
$\epsilon_{\text B} \ll \epsilon_{\text F}$, and we recover the
well-known BCS result with strongly overlapping Cooper pairs. In
this limit we have the chemical potential
$\mu_0\simeq\epsilon_{\text F}$ and the gap function
$\Delta_0\ll\epsilon_{\text F}$. For the opposite limit of very
strong attraction (or low particle density), we have a deep
two-body bound state with $\epsilon_{\text B} \gg \epsilon_{\text
F}$, and the system is in the BEC region with composite bosons. In
this limit the chemical potential takes $\mu_0 \simeq
-\epsilon_{\text B}/2$. It should be kept in mind that in the
local pair regime ($\mu_0<0$) the fermion excitation gap
$E_{\text{gap}}$ in the quasi-particle excitation spectrum is not
$\Delta_0$ (as in the case $\mu_0> 0$) but rather
$\sqrt{\mu_0^{2}+\Delta_0^{2}}$.

In ultracold Fermi gas experiments, a quasi-2D Fermi gas can be
realized by arranging a one-dimensional optical lattice along the
axial ($z$) direction and a weak harmonic trapping potential in
the radial ($x$-$y$) plane, such that fermions are strongly
confined along the $z$ direction and form a series of
pancake-shaped clouds~\cite{2Dexp,zhang}. Each such cloud can be
considered as a quasi-2D Fermi gas when the axial confinement is
strong enough to turn off inter-cloud tunnelling. The strong
anisotropy of the trapping potentials, namely $\omega_{z}\gg
\omega$ where $\omega_z$($\omega$) is the axial(radial) frequency,
allows us to use an effective 2D Hamiltonian to deal with the
radial degrees of freedom\cite{zhang}. The effective binding
energy $\epsilon_{\text B}=\hbar\omega_z\exp{[4\pi
a_z^2/U_{p}^{\text{eff}}(a_s,a_z)]}$ is related to the energy
scale $\hbar\omega_z$ and the 3D s-wave scattering length $a_s$,
where $a_z$ is defined as $a_z=\sqrt{\hbar/(m\omega_z)}$ and the
quantity $U_{p}^{\text{eff}}(a_s,a_z)$ defined in \cite{zhang}
carries the dependence on 3D scattering length. By adjusting the
3D scattering length $a_s$ and/or the axial frequency $\omega_z$,
a quasi-2D BCS-BEC crossover can be realized.

\section {Equation of State}
We now turn on the Zeeman splitting $h\neq 0$. To determine the
superfluid phase diagrams and calculate the density profile for
imbalanced Fermi gas in a harmonic trap, we first establish the
equations of state~(EOS) for various phases in grand canonical
ensemble\cite{Sheehy}. In the BCS-Leggett mean field theory, the
pressure ${\cal P}=-\Omega$ as a function of $\mu$ and $h$ can be
evaluated as\cite{Sheehy}
\begin{eqnarray}
{\cal P}(\mu,h)&=&c\int_0^\infty dz\bigg[E_z-z+\mu-\frac{\Delta^2}{2z+\epsilon_{\text{B}}}\nonumber\\
&-&(E_z-h)\Theta(h-E_z)\bigg]
\end{eqnarray}
with $ E_z=\sqrt{(z-\mu)^2+\Delta^2}$ and $c=M/(2\pi)$. We have
set $h>0$ without loss of generality. The superfluid order
parameter
$\Delta(\mu,h)=-U\langle\psi_{\downarrow}^{\phantom{\dag}}\psi_{\uparrow}^{\phantom{\dag}}\rangle$
is determined self-consistently from the gap equation
\begin{equation}
\Delta\int_0^\infty
dz\left[\frac{1}{2z+\epsilon_{\text{B}}}-\frac{\Theta(E_z-h)}{2E_z}\right]=0.
\end{equation}
The step function $\Theta(x)$ in this paper is defined as
$\Theta(x)=0$ for $x<0$ and $\Theta(x)=1$ for $x>0$.

Unlike the 3D case\cite{Sheehy}, the EOS in 2D can be analytically
obtained. At fixed $\mu$ and $h$, we have three possible phases:
the unpolarized superfluid phase (SF), the polarized normal phase
(N) and the polarized superfluid phase or Sarma phase
(S)~\cite{Sarma}. The phase SF corresponds to the solution
$\Delta(\mu)=\sqrt{\epsilon_{\text B}(\epsilon_{\text B}+2\mu)}$
in the region $h<E_{\text
g}=\sqrt{\mu^2\Theta(-\mu)+\Delta^2(\mu)}$, and the pressure can
be evaluated as
\begin{eqnarray}
{\cal P}_{\text{SF}}(\mu)=c\left(\mu+\frac{\epsilon_{\text
B}}{2}\right)^2
\end{eqnarray}
which does not depend explicitly on $h$. The total number density
$n=n_\uparrow+n_\downarrow$ and the magnetization
$m=n_\uparrow-n_\downarrow$ can be expressed as
\begin{eqnarray}
n_{\text{SF}}(\mu) &=& 2c\left(\mu+\frac{\epsilon_{\text
B}}{2}\right),\ \ \ m_{\text{SF}}(\mu)=0. \label{EOSSF}
\end{eqnarray}
The polarized superfluid phase or Sarma phase~(S) corresponds to
the solution in the region $h>E_{\text g}$. This phase can be
ruled out from the positive secondary derivative~\cite{guba}
\begin{equation}
\frac{\partial^2{\cal P}}{\partial \Delta^2}\Big|_{\text
S}=c\left[\frac{h(1+\Theta(\mu))}{\sqrt{h^2-\Delta^2}}-\frac{\mu\Theta(\mu)}{\sqrt{\mu^2+\Delta^2}}-1\right]>0,\label{Sarma}
\end{equation}
which means that the Sarma phase is always unstable for any
coupling in 2D.

The polarized normal phase corresponds to the solution $\Delta=0$.
The pressure takes the form of non-interacting Fermi gas,
\begin{eqnarray}
{\cal P}_{\text
N}(\mu,h)=\frac{c}{2}\left[(\mu-h)^2\Theta(\mu-h)+(\mu+h)^2\Theta(\mu+h)\right],\nonumber\\
\end{eqnarray}
where the case $\mu+h<0$ corresponds to the vacuum without atoms.
For $\mu+h>0$, the total number density and the magnetization read
\begin{eqnarray}
n_{\text N}(\mu,h)&=&2c\mu\Theta\left(\mu-h\right)
+c(\mu+h)\Theta\left(h-\mu\right),\nonumber\\
m_{\text N}(\mu,h)&=&2ch\Theta\left(\mu-h\right)
+c(\mu+h)\Theta\left(h-\mu\right).\label{EOSN}
\end{eqnarray}
The cases $\mu>h$ and $\mu<h$ correspond to the partially
polarized (N$_{\text{PP}}$) and fully polarized (N$_{\text{FP}}$)
normal phases respectively. Since we treat the superfluid and
normal phase in mean field approximation, the normal phase is
considered as a non-interacting gas. In fully polarized case, this
is correct since only s-wave interaction is considered. However,
in partially polarized case, the interaction may be important in
some coupling parameter region, like the finding around the
unitary region in 3D~\cite{unitary,MC1,MC2}. Including
fluctuations, which can not be treated analytically even in 2D, is
necessary for a more realistic study.

Since the polarized superfluid phase is always located at the
maximum of the thermodynamic potential, there exists at fixed
$\mu$ a first order quantum phase transition from the SF phase to
the normal phase when the Zeeman field $h$ increases. The critical
value $h_c$ is determined by the condition ${\cal
P}_{\text{SF}}(\mu)={\cal P}_{\text{N}}(\mu, h_c)$. The analytical
expression for $h_c$ can be written as
\begin{eqnarray}
h_c(\mu)&=&\sqrt{\epsilon_{\text B}\left(\mu+\frac{\epsilon_{\text
B}}{4}\right)}\Theta(\mu-h_0)\nonumber\\
&+&\left[(\sqrt{2}-1)\mu+\frac{\epsilon_{\text
B}}{\sqrt{2}}\right]\Theta(h_0-\mu).
\end{eqnarray}
Equivalently, for a given $h$, SF-N phase transition happens when
the chemical potential $\mu$ becomes less than the critical value
\begin{eqnarray}
\mu_c(h)&=&\left(\frac{h^2}{\epsilon_{\text
B}}-\frac{\epsilon_{\text
B}}{4}\right)\Theta\left(h-h_0\right)\nonumber\\
&+&\frac{\sqrt{2}h-\epsilon_{\text
B}}{2-\sqrt{2}}\Theta\left(h_0-h\right), \label{UC}
\end{eqnarray}
where $h_0=(\sqrt{2}+1)\epsilon_{\text B}/2$ is determined by the
equation $h_0=\mu_c(h_0)$. We can easily prove that
$h>h_0$($h<h_0$) is equivalent to the condition
$\mu_c>h$($\mu_c<h$).

\begin{figure}[!htb]
\begin{center}
\includegraphics[width=8cm]{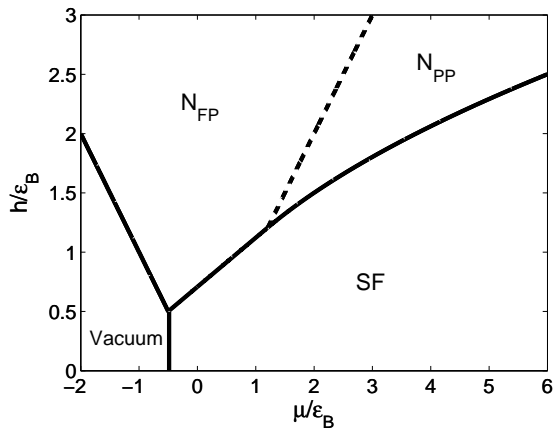}
\caption{The grand canonical phase diagram in the $\mu-h$ plane.
$\mu$ and $h$ are scaled by the binding energy $\epsilon_{\text
B}$. \label{fig1}}
\end{center}
\end{figure}

The grand canonical phase diagram in the $\mu-h$ plane is shown in
Fig.\ref{fig1}. The analytical expressions for the phase
boundaries can be obtained from the above expression for $h_c$.
The SF phase, N$_{\text{FP}}$ phase and the vacuum meet at the
point $(\mu,h)=(-\epsilon_{\text B}/2,\epsilon_{\text B}/2)$,
while the three phases SF, N$_{\text{PP}}$ and N$_{\text{FP}}$
meet at $(\mu,h)=((\sqrt{2}+1)\epsilon_{\text
B}/2,(\sqrt{2}+1)\epsilon_{\text B}/2)$. The grand canonical phase
diagram is of great help for us to understand the density profile
in a harmonic trap.

\section {Homogeneous Fermi Gas}
In this section we determine the phase diagram of the homogeneous
system. Since the total atom number $N=N_\uparrow+N_\downarrow$ or
equivalently the total atom density $n$ is fixed, the chemical
potential $\mu$ is not a free variable in the canonical ensemble
and should be determined by the number conservation. One may
distinguish two different cases: (1)The Zeeman field $h$ can be
experimentally adjusted by using Raman detuning\cite{liu}; (2)The
atom number for each species, $N_\uparrow$ and $N_\downarrow$, can
be adjusted\cite{exp1,exp2}. Since the phase structure should be
essentially independent of the ensemble we choose, we firstly
discuss the phase diagram using $h$ as tunable parameter, and then
translate it into the case where the global polarization
$P=(N_{\uparrow}-N_{\downarrow})/(N_{\uparrow}+N_{\downarrow})$ is
directly adjusted.

\subsection {Critical Zeeman Fields}
We now consider the problem: When does the superfluidity disappear
when a Zeeman splitting $h$ is turned on? The total density $n$
and the Zeeman field $h$ are thermodynamic variables, and the free
energy of the system should be defined as ${\cal F}(n,h)=\mu
n-{\cal P}$. Since $n=M\epsilon_{\text F}/\pi$ is fixed, we will
write ${\cal F}(n,h)={\cal F}(h)$. At nonzero Zeeman field $h$,
the solutions of the coupled gap and number equations
corresponding to the above three homogeneous bulk phases can be
analytically solved:

\noindent {\bf I}. $\Delta_{\text{SF}}(h)=\Delta_0$ and
$\mu_{\text{SF}}(h)=\mu_0$ in the phase SF. The solution exists in
the region $0<h<\Delta_0$ for $\eta<2$ or $0<h<\epsilon_{\text
F}+\epsilon_{\text B}/2$ for $\eta>2$.

\noindent {\bf II}. $\Delta_{\text N}(h)=0$ and $\mu_{\text
N}(h)=\epsilon_{\text F}\Theta(\epsilon_{\text
F}-h)+(2\epsilon_{\text F}-h)\Theta(h-\epsilon_{\text F})$ in the
phase N. The first and second term correspond, respectively, to
partially and fully polarized normal phase.

\noindent {\bf III}.
$\Delta_{\text{S}}(h)=\sqrt{\Delta_0(2h-\Delta_0)}$ and
$\mu_{\text S}(h)=\epsilon_{\text F}-
\Delta_{\text{S}}^2(h)/(4\epsilon_{\text F})$ in the region of
$\Delta_0/2<h<\Delta_0$ and $\eta < 2$ and
$\Delta_{\text{S}}(h)=\sqrt{\epsilon_{\text B}(2h-\epsilon_{\text
B})}$ and $\mu_{\text S}(h)=2\epsilon_{\text F}- h$ in the region
of $\epsilon_{\text B}/2<h<\epsilon_{\text F}+\epsilon_{\text
B}/2$ and $\eta > 2$ in the phase S. There are two gapless Fermi
surfaces at $\eta < 2$ and only one gapless Fermi surface at $\eta
> 2$.

The polarized superfluid phase or the Sarma phase, which is a
gapless superfluid, is again an unstable state at any coupling,
directly from the reentrance phenomenon (three solutions of
$\Delta$ at fixed $h$), in contrast to the case in 3D where it
becomes the stable ground state in the strong coupling BEC
region~\cite{Sheehy,MFwork}. This is an important difference of
the Fermi surface topology and the stability condition between 3D
and 2D cases\cite{guba}. Explicitly, the free energy (density)
${\cal F}(h)=\mu(h) n-{\cal P}(\mu(h),h)$ in the three homogeneous
bulk phases reads
\begin{eqnarray}
{\cal F}_{\text{SF}}(h)&=&c(\epsilon_{\text F}^2-\epsilon_{\text
F}\epsilon_{\text B}),\nonumber\\
{\cal F}_{\text{N}}(h)&=&c\big[(\epsilon_{\text
F}^2-h^2)\Theta(\epsilon_{\text
F}-h)\nonumber\\
&+&2(\epsilon_{\text F}^2-\epsilon_{\text
F}h)\Theta(h-\epsilon_{\text
F})\big],\nonumber\\
{\cal F}_{\text{S}}(h)&=&c[2(\epsilon_{\text F}^2-\epsilon_{\text
F}h)+h^2+(h-\epsilon_{\text B}/2)^2]\Theta(\eta-2)\nonumber\\
&+&c[(\epsilon_{\text F}^2-\epsilon_{\text F}\epsilon_{\text
B})+(\Delta_0-h)^2]\Theta(2-\eta).
\end{eqnarray}
It is easy to see that the polarized superfluid phase has always
higher free energy. If there exist no other possible phases, a
first order quantum phase transition from the phase SF to the
phase N will occur at a critical Zeeman field $h_{c}$ determined
by ${\cal F}_{\text{SF}}(h_c)={\cal F}_{\text{N}}(h_c)$. We find
$h_c=\sqrt{\eta}\epsilon_{\text F}=\Delta_0/\sqrt{2}$ for $\eta<1$
and $h_c=\frac{1}{2}(1+\eta)\epsilon_{\text F}$ for $\eta>1$. It
is interesting to note that the relation $h_c=\Delta_0/\sqrt{2}$
at $\eta<1$ is only an approximate result at weak coupling in
3D~\cite{CC,Sarma}.

If there are only the two bulk phases SF and N, we have only one
CC limit at which the first order phase transition occurs, and the
experimentally observed phase separation (PS) will be hidden in
the $\eta$-$h$ phase diagram. However, since the total atom
density $n$ is fixed, unlike the grand canonical ensemble, we
should consider possible mixed phases constructed via the Gibbs
phase equilibrium condition. Here we will neglect the interfacial
energy\cite{surface}, since for a macroscopic phase separation,
this energy contribution is subdominant in the thermodynamic
limit. From equation (\ref{Sarma}), the only possibility is the
SF-N mixed phase. When the phase separation is favored in a region
$h_{c1}<h<h_{c2}$, the chemical potential $\mu_{\text{PS}}$ is
different from $\mu_{\text{SF}}$ and $\mu_{\text{N}}$, it should
be determined by the phase equilibrium condition ${\cal
P}_{\text{SF}}(\mu)={\cal P}_{\text N}(\mu,h)$, which leads to
\begin{eqnarray}
\mu_{\text{PS}}(h)&=&\left(\frac{h^2}{\epsilon_{\text
B}}-\frac{\epsilon_{\text
B}}{4}\right)\Theta\left(h-h_0\right)\nonumber\\
&+&\frac{\sqrt{2}h-\epsilon_{\text
B}}{2-\sqrt{2}}\Theta\left(h_0-h\right), \label{UPS}
\end{eqnarray}
where $h_0=(\sqrt{2}+1)\epsilon_{\text B}/2$ satisfies the
equation $h_0=\mu_{\text{PS}}(h_0)$. Since the chemical potential
$\mu_{\text PS}$ is determined by the condition ${\cal
P}_{\text{SF}}(\mu)={\cal P}_{\text N}(\mu,h)$, it is equivalent
to the critical chemical potential $\mu_c(h)$ in the grand
canonical ensemble. The cases $h>h_0$ and $h<h_0$ indicate,
respectively, the mixed phases with partially polarized normal
bubbles (SF-N$_{\text{PP}}$) and fully polarized normal bubbles
(SF-N$_{\text{FP}}$). The volume fractions of the phases SF and N
in the phase separation, denoted by $x$ and $1-x$ respectively,
are determined by the number conservation,
$n=x(h)n_{\text{SF}}(\mu_{\text{PS}},h)+[1-x(h)]n_{\text
N}(\mu_{\text{PS}},h)$. Using the expressions (\ref{EOSSF}),
(\ref{EOSN}) and (\ref{UPS}) for $\mu_{\text{PS}}, n_{\text N}$
and $n_{\text{SF}}$, we find
\begin{eqnarray}
x(h)&=&2\left(\frac{\epsilon_{\text{F}}}{\epsilon_{\text{B}}}+\frac{1}{4}-\frac{h^2}{\epsilon_{\text{B}}^2}\right)\Theta\left(h-h_0\right)\nonumber\\
&+&\left(\frac{2\sqrt{2}\epsilon_{\text F}}{2h-\epsilon_{\text
B}}-\sqrt{2}-1\right)\Theta\left(h_0-h\right).
\end{eqnarray}

We now determine the region of the mixed phase, i.e., the lower
and upper critical fields $h_{c1}$ and $h_{c2}$~\cite{Sheehy}. In
the grand canonical ensemble with fixed chemical potential $\mu$,
we have only one critical field $h_c(\mu)$ determined by the
condition ${\cal P}_{\text{SF}}(\mu,h)={\cal P}_{\text N}(\mu,h)$,
and the signal of SF-N phase separation is denoted by the first
order phase transition line in the $\mu-h$ phase diagram. In the
standard BCS-BEC crossover problem, the total atom number $N$
rather than the chemical potential $\mu$ is fixed, and the CC
limit splits into two values $h_{c1}=h_c(\mu_{\text{SF}})$ and
$h_{c2}=h_c(\mu_{\text N})$~\cite{Sheehy}. The mixed phase links
continuously the phases SF and N with
$\mu_{\text{PS}}=\mu_{\text{SF}}$ at $h=h_{c1}$ and
$\mu_{\text{PS}}=\mu_{\text{N}}$ at $h=h_{c2}$ and ensures $0\leq
x\leq 1$ with $x(h_{c1})=1$ and $x(h_{c2})=0$. The critical fields
$h_{c1}$ and $h_{c2}$ are explicitly given by
\begin{eqnarray}
h_{c1}&=&\epsilon_{\text{F}}\sqrt{\eta\left(1-\frac{\eta}{4}\right)}\Theta(\eta_1-\eta),\nonumber\\
&+&\epsilon_{\text{F}}\left(\sqrt{2}-1+\frac{\eta}{2}\right)\Theta(\eta-\eta_1),\nonumber\\
h_{c2}&=&\epsilon_{\text{F}}\sqrt{\eta\left(1+\frac{\eta}{4}\right)}\Theta(\eta_2-\eta),\nonumber\\
&+&\epsilon_{\text{F}}\left(2-\sqrt{2}+\frac{\eta}{2}\right)\Theta(\eta-\eta_2),
\end{eqnarray}
where $\eta_1=2-\sqrt{2}\simeq0.586$ and
$\eta_2=2(\sqrt{2}-1)\simeq0.828$ are determined by
$\mu_{\text{SF}}(h_0)=h_0$ and $\mu_{\text N}(h_0)=h_0$,
respectively. There is always the relation $h_{c1}<h_{c}<h_{c2}$,
and the splitting disappears in the weak coupling limit
$\eta\rightarrow 0$ which recovers the well known result shown in
\cite{CC,Sarma}. On the other hand, the splitting keeps as a
constant $(3-2\sqrt{2})\epsilon_{\text F}\simeq
0.172\epsilon_{\text F}$ at strong coupling $\eta>\eta_2$.

The final step is to prove that the SF-N mixed phase has the
lowest free energy in the region $h_{c1}<h<h_{c2}$. Using the
analytical expressions for $x(h)$ and $\mu_{\text{PS}}(h)$ as well
as the EOS for the phases SF and N, we can evaluate the free
energy in the mixed phase defined by ${\cal
F}_{\text{PS}}(h)=\mu_{\text{PS}}n-x(h){\cal
P}_{\text{SF}}(\mu_{\text{PS}},h)-[1-x(h)]{\cal
P}_{\text{N}}(\mu_{\text{PS}},h)$. The difference between ${\cal
F}_{\text{PS}}$ and ${\cal F}_{\text{SF}}$ and between ${\cal
F}_{\text{PS}}$ and ${\cal F}_{\text{N}}$ can be explicitly
expressed as
\begin{eqnarray}
&&{\cal F}_{\text{PS}}(h)-{\cal F}_{\text{SF}}(h)=-\epsilon_{\text
B}^{-2}c(h^2-h_{c1}^2)^2\Theta(h-h_0)\nonumber\\
&&\ \ \ \ \ \ \ \ \ \ \ \ \ \ \ \ \ -(\sqrt{2}+1)^2c(h-h_{c1})^2\Theta(h_0-h),\nonumber\\
&&{\cal F}_{\text{PS}}(h)-{\cal F}_{\text{N}}(h)=-\epsilon_{\text
B}^{-2}c(h^2-h_{c2}^2)^2\Theta(h-h_0)\nonumber\\
&&\ \ \ \ \ \ \ \ \ \ \ \ \ \ \ \ \
-(\sqrt{2}+1)^2c(h-h_{c2})^2\Theta(h_0-h).
\end{eqnarray}
The above expressions show explicitly that the mixed phase has
really the lowest free energy in the region $h_{c1}<h<h_{c2}$.
While in 3D the conclusion that the mixed phase corresponds to the
lowest free energy is analytically proven in the weak coupling
limit~\cite{BCR}, our result here in 2D is for any coupling.

The SF-N mixed phase has a nonzero global polarization $P$ since
the normal bubble is polarized. From the definition
$P=(N_{\uparrow}-N_{\downarrow})/(N_{\uparrow}+N_{\downarrow})$ we
find
\begin{eqnarray}
P(h)=[1-x(h)]\frac{m_{\text N}(\mu_{\text{PS}},h)}{n}.
\end{eqnarray}
Using the expression for $x(h)$ and $\mu_{\text{PS}}$, we have
\begin{eqnarray}
P(h)&=&\frac{2h(h^2-h_{c1}^2)}{\epsilon_{\text F}\epsilon_{\text
B}^2}\Theta(h-h_0)\nonumber\\
&+&\frac{(\sqrt{2}+1)^2(h-h_{c1})}{\epsilon_{\text
F}}\Theta(h_0-h).
\end{eqnarray}
The global polarization is zero at $h=h_{c1}$ and then increases
with $h$.

\subsection {Critical Polarization}
Finally, we convert the above result into the one where both
$N_\uparrow$ and $N_\downarrow$ are fixed and the exchange between
particles in states $\uparrow$ and $\downarrow$ is forbidden,
corresponding to recent experiments on ultracold Fermi gas with
population imbalance\cite{exp1,exp2}. The free energy density in
this case should be defined as ${\cal
F}(n_\uparrow,n_\downarrow)=\mu_\uparrow n_\uparrow+\mu_\downarrow
n_\downarrow-{\cal P}$ or ${\cal F}(n,m)=\mu n+ mh-{\cal P}$. The
possible phases with nonzero global polarization $P$ are the
normal, Sarma and SF-N mixed phases. Since the phase structure
should be essentially independent of the ensemble we
choose\cite{Sheehy}, we do not need to compare again the free
energies of the three phases\cite{BCR}. Since $P(h_{c1})=0$ and
$P(h)$ increases with $h$, we conclude that the ground state is
the unpolarized superfluid at $P=0$, and SF-N phase separation
becomes energetically favored for $0<P<P_c$. The critical
polarization $P_c$ where the superfluid bubble disappears
completely is the global polarization at $h_{c2}$,
\begin{eqnarray}
P_c&=&P(h_{c2})=\frac{h_{c2}}{\epsilon_{\text{F}}}\Theta(\eta_2-\eta)+\Theta(\eta-\eta_2)\nonumber\\
&=&\sqrt{\eta(1+\frac{\eta}{4})}\Theta(\eta_2-\eta)+\Theta(\eta-\eta_2).
\end{eqnarray}
The critical polarization increases from $P_c=0$ at $\eta=0$ to
$P_c=1$ at $\eta=\eta_2$ and then keeps as a constant $P_c=1$ at
strong coupling $\eta>\eta_2$

\begin{figure}[!htb]
\begin{center}
\includegraphics[width=8cm]{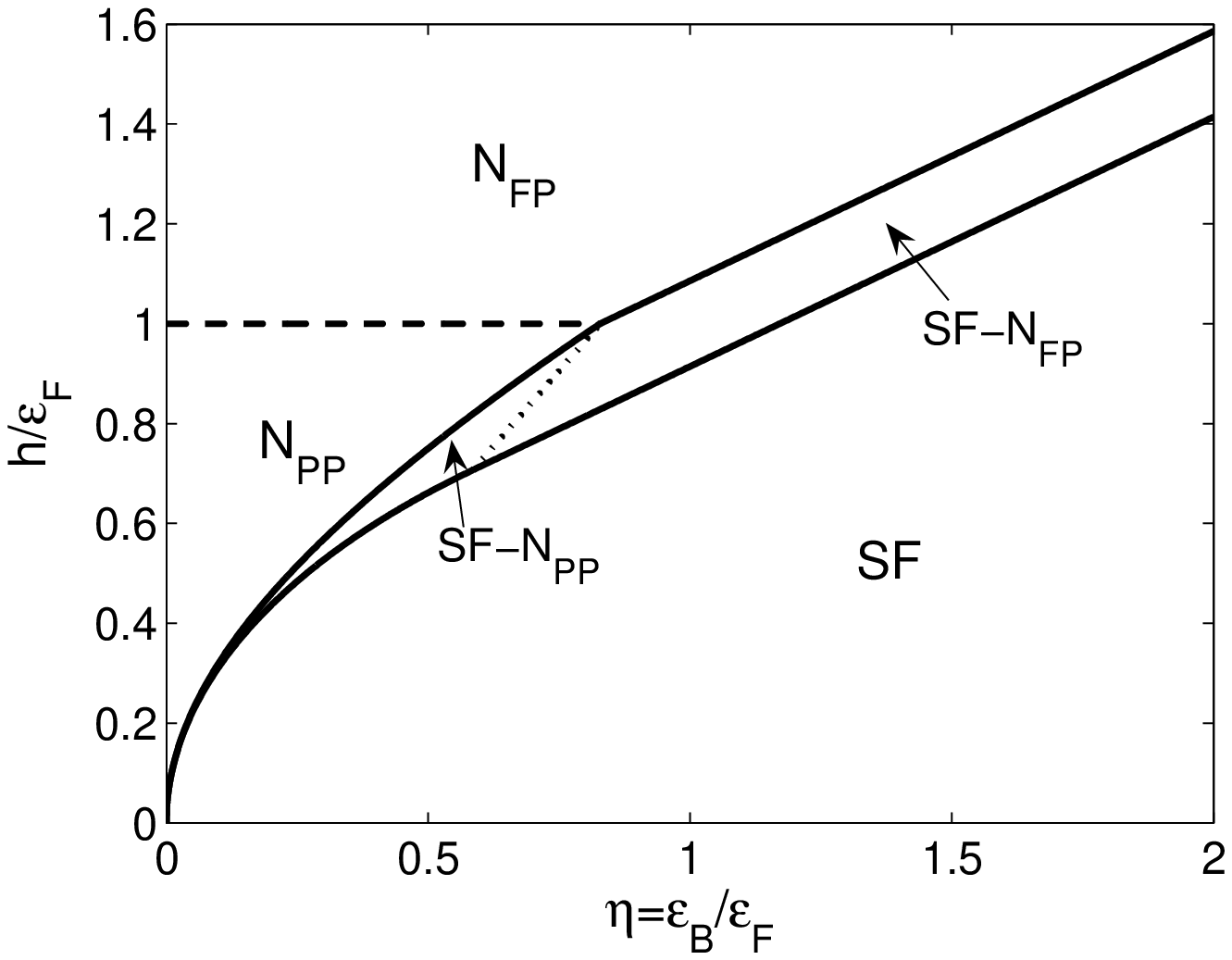}
\includegraphics[width=8cm]{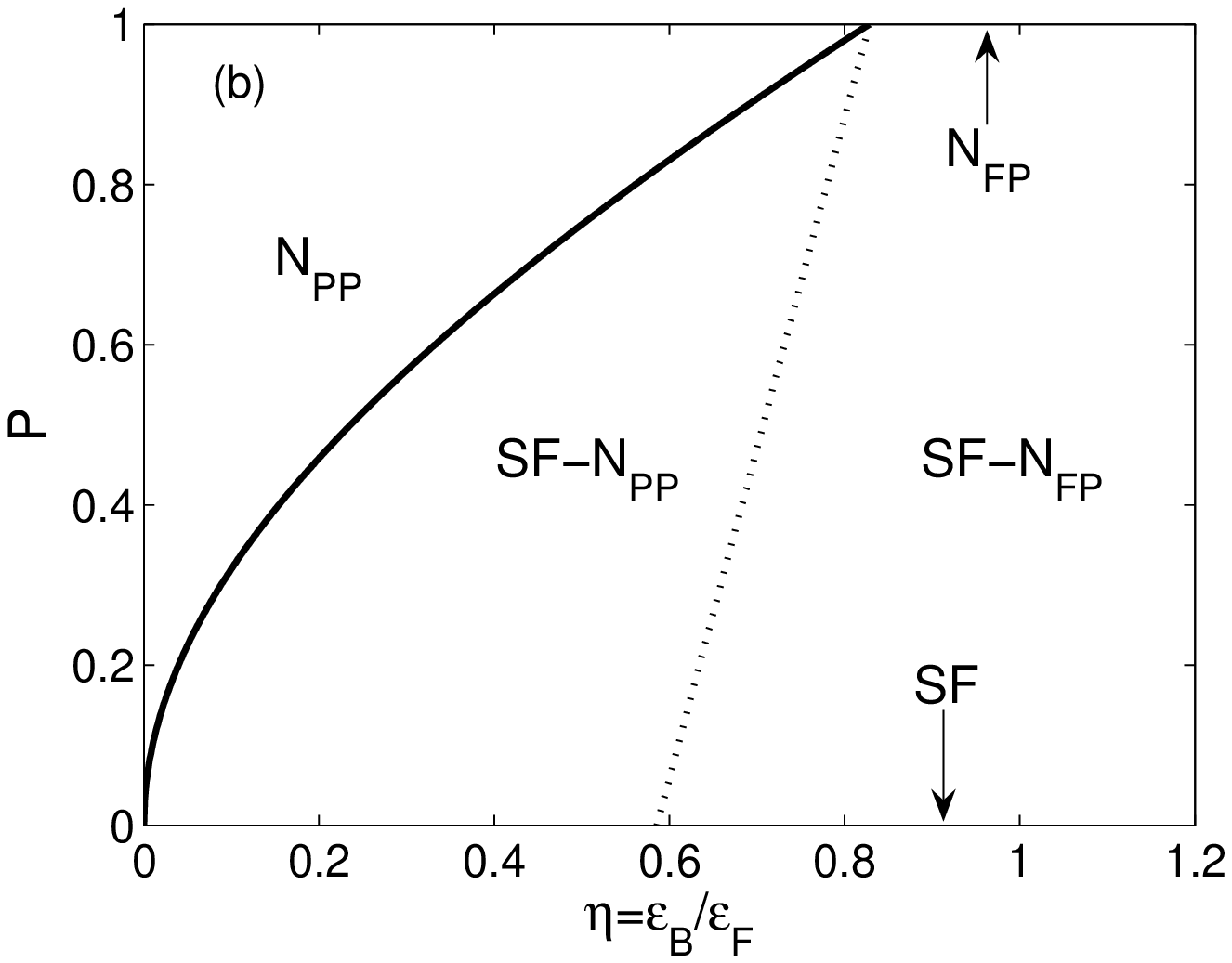}
\caption{The phase diagrams in the planes $\eta-h$ (upper panel)
and $\eta-P$ (lower panel). $h$ is scaled by the Fermi energy. SF
means unpolarized superfluid, $N_{\text{PP}}$ and $N_{\text{FP}}$
indicate the partially and fully polarized normal phases, and
SF-N$_{\text {PP}}$ and SF-N$_{\text {FP}}$ are the mixed phases
of superfluid and normal gas with $N_{\text{PP}}$ and
$N_{\text{FP}}$. \label{fig2}}
\end{center}
\end{figure}

\subsection {Phase Diagrams}
Fig.\ref{fig2} summarizes the above analytical results. The phase
diagram in the $\eta-h$ plane is shown in the upper panel. The
partially and fully polarized normal phases $N_{\text{PP}}$ and
$N_{\text{FP}}$ are separated by the dashed line
$h/\epsilon_{\text F}=1$ which ends at $\eta=\eta_2$. The two
solid lines indicate the lower and upper critical Zeeman fields
$h_{c1}$ and $h_{c2}$ with the two phase separations PS-I and
PS-II in between. PS-I (SF-N$_{\text {PP}}$)and PS-II
(SF-N$_{\text {FP}}$) are the mixed phases of superfluid and
normal gas with $N_{\text{PP}}$ and $N_{\text{FP}}$, and they are
separated by the dotted line $h/\epsilon_{\text
F}=(\sqrt{2}+1)\eta/2$ starting at $\eta=\eta_1$ and ending at
$\eta=\eta_2$. The phase diagram in the $\eta-h$ plane can be
easily converted into the one in the $\eta-P$ plane shown in the
lower panel, by taking the fact $P(h_{c1})=0$ and $P(h_{c2})=P_c$.
The critical polarization $P_c=\sqrt{\eta(1+\eta/4)}$ (solid line)
increases from $P_c=0$ at $\eta=0$ to $P_c=1$ at $\eta=\eta_2$ and
then keeps as a constant $P_c=1$ for $\eta>\eta_2$. The phases SF
and $N_{\text{FP}}$ are now located at $P=0$ and $P=1$
respectively. The dotted line which separates PS-I from PS-II
becomes $P=(4+3\sqrt{2})\eta/2-(\sqrt{2}+1)$ in the $\eta-P$
plane.

The above analytical results show that, to correctly calculate the
critical polarization $P_c$ and the phase diagrams, one should
treat the mixed phase carefully~\cite{Sheehy}. Some other methods
taken in literatures may lead to quantitatively incorrect results.
For instance, the method of stability analysis will result in an
incorrect critical polarization (see also the comments in
\cite{comment}). With this method, one first solve the mean field
gap and number equations for the Sarma phase and then analyze the
stability of this phase. If it is applied to the 2D system, the
critical polarization becomes~\cite{tem}
\begin{eqnarray}
\label{eqq}
P_c=\frac{\Delta_0}{2\epsilon_{\text
F}}=\sqrt{\frac{\eta}{2}},
\end{eqnarray}
which is the maximum polarization of the unstable Sarma phase and
deviates significantly from our result
$P_c=\sqrt{\eta(1+\eta/4)}$. Especially, our critical polarization
grows up to unity at $\eta\simeq0.828$, but the result (\ref{eqq})
becomes unity at $\eta=2$. On the other hand, if one takes only
the phases SF and N into account but neglect the phase separation,
there will be only one critical field $h_c$ where the polarization
jumps from $0$ to $h_c/\epsilon_{\text F}=\sqrt{\eta}$.

\section {Bound State in Polarized Normal Phase}
In recent experiment on highly polarized normal phase in 3D
unitary Fermi gas\cite{exp5}, it is found that while the
superfluidity disappears completely, full pairing of minority
atoms always exists, which indicates that the fermion pairing may
be easy to occur in the presence of polarization. It is well known
that the bound state in 2D can form at arbitrary small attractive
interaction\cite{QM}, which is quite different to the 3D case. It
is natural to ask: Do the di-fermion bound states exist above the
upper critical field $h_{c2}$ or critical polarization $P_c$?

In this section, we study the spectrum of bound states in the
highly polarized normal phase. In the Green function method, the
energy $\omega$ of the bound states with zero total momentum in
this case is determined through the equation~\cite{2Dreview}
\begin{equation}
\int_{0}^{\infty} dz \left[\frac{1}{2z+\epsilon_{\text B}} -
\frac{1-\Theta(\mu_\uparrow-z)-\Theta(\mu_\downarrow-z)}{2z-2\mu-\omega}\right]=0.
\end{equation}
In the vacuum with $\mu=h=0$, it self-consistently gives the
solution $\omega=-\epsilon_{\text B}$. In general case with medium
effect, the bound states can survive when the above equation has
real solution of $\omega$.

The integration in the above equation can be analytically worked
out, and finally we obtain
\begin{equation}
\ln\frac{\omega+2\mu}{-\epsilon_{\text
B}}+\Theta(\mu-h)\ln\frac{\omega+2h}{\omega+2\mu}+\Theta(\mu+h)\ln\frac{\omega-2h}{\omega+2\mu}=0.
\end{equation}
In the partially polarized normal phase, we have
$\mu=\epsilon_{\text F}$, the spectrum equation becomes
\begin{equation}
\omega^2+\epsilon_{\text B}\omega+2\epsilon_{\text
F}\epsilon_{\text B}-4h^2=0
\end{equation}
which has real solutions
\begin{equation}
\omega=-\frac{1}{2}\left(\epsilon_{\text
B}\pm\sqrt{J(h)}\right)\label{ph}
\end{equation}
for
\begin{equation}
J(h)=\epsilon_{\text B}^2+16h^2-8\epsilon_{\text B}\epsilon_{\text
F}>0.
\end{equation}
In the fully polarized normal phase, one finds that the spectrum
equation directly gives a real solution $\omega=2h-\epsilon_{\text
B}$. However, this solution is unphysical since we always have
$\omega>0$. This can be well understood when we consider the fact
that there exist only $\uparrow$ particles in this phase.

\begin{figure}[!htb]
\begin{center}
\includegraphics[width=8cm]{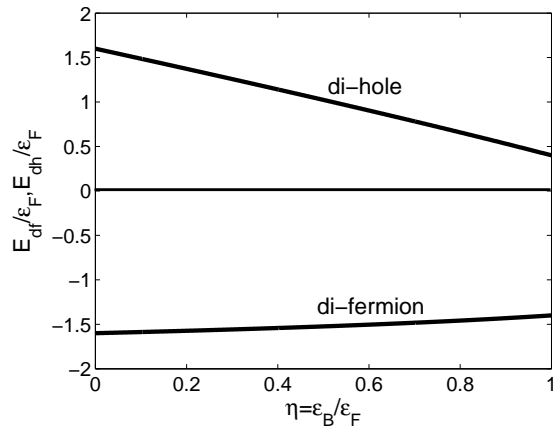}
\caption{The excitation gaps for the di-fermions and di-holes as a
function of $\eta$ at fixed polarization $P=0.8$. The curves are
meaningful only for $\eta<0.56$, since the ground state is not a
normal phase for $\eta>0.56$. \label{fig3}}
\end{center}
\end{figure}

In the balanced normal phase with $h=0$ (note that the true ground
state in this case is the superfluid phase), the bound states
remain stable only at strong enough coupling $\eta>8$ or
equivalently low enough density $\epsilon_{\text
F}<\epsilon_{\text B}/8$~\cite{2Dreview}, which indicates that the
Fermi sea or medium effect disfavors the formation of bound
states. One may simply think that, the presence of a Zeeman
splitting will further destroy the bound states. However, this is
not true. Since the condition $J(h)>0$ is easier to be satisfied
at $h\neq 0$, the bound states in highly polarized normal phase
are easier to survive than in the balanced Fermi sea. In the whole
partially polarized normal phase which exists in the region
$0<\eta<\eta_2$ in Fig.\ref{fig2}, we have $\epsilon_{\text
F}\sqrt{\eta(1-\eta/4)}<h<\epsilon_{\text F}$. Analyzing the
condition $J(h)>0$, we conclude the bound states can exist in the
whole N$_{\text{PP}}$ phase in Fig.\ref{fig2}. Especially, they
can survive at high polarization even in the weak coupling limit,
as pointed out in~\cite{fum} in the case of 3D.

There exist two real solutions for $\omega$ in the N$_{\text{PP}}$
phase. The negative~($\omega<0$) and positive~($\omega>0$)
solutions in (\ref{ph}) correspond to the excitation gaps
$E_{\text{df}}$ and $E_{\text{dh}}$ for the di-fermions(df) and
di-holes(dh) respectively\cite{fum},
\begin{eqnarray}
E_{\text{df}}&=&-\frac{\epsilon_{\text
F}}{2}\left(\eta+\sqrt{\eta^2-8\eta+16P^2}\right),\nonumber\\
E_{\text{dh}}&=&\frac{\epsilon_{\text
F}}{2}\left(\sqrt{\eta^2-8\eta+16P^2}-\eta\right).
\end{eqnarray}
In Fig.\ref{fig3}, we plot the excitation gaps for di-fermions and
di-holes at a fixed polarization $P=0.8$. We found that the
symmetry in the spectrum($E_{\text{df}}=-E_{\text{dh}}$) holds
only at weak coupling.

\section {Density Profile in a Harmonic Trap}
We have determined the phase diagram for homogeneous system.
However, the phase structure can not be directly examined in
ultracold Fermi gas experiments, due to the effect of the external
harmonic trap. To justify the theoretical prediction for
homogeneous system, one should calculate the corresponding phase
diagram and the density profile in the presence of an external
trap potential, using the same equation of state. In this section,
we will calculate analytically the density profile of an imbalance
Fermi gas in a 2D isotropic harmonic trap potential
$V(r)=\frac{1}{2}M\omega^2r^2$. The frequency $\omega$ here is
different from the energy of the bound state defined in Section V.

The effect of a harmonic trap can be treated in the local density
approximation(LDA). In the frame of LDA, the system is
approximately taken to be uniform but with a local chemical
potential given by
\begin{eqnarray}
\mu(r)=\mu_0-\frac{1}{2}M\omega^2r^2,
\end{eqnarray}
where $\mu_0$ is the chemical potential at the center of the trap
and is the true chemical potential(a Lagrangian multiplier) still
enforcing the total atom number $N=N_\uparrow+N_\downarrow$. Since
$N_\uparrow$ and $N_\downarrow$ are both conserved, the
spatially-varying spin-up and spin-down local chemical potentials
can be expressed as $\mu_\uparrow(r)=\mu(r)+h$ and
$\mu_\downarrow(r)=\mu(r)-h$ in terms of the averaged chemical
potential $\mu(r)$ and Zeeman field $h$.

To calculate the density profile, namely the atom density of the
spin-up and spin-down states as a function of the radius $r$,
$n_\uparrow(r)$ and $n_\downarrow(r)$, or equivalently the total
density $n(r)=n_\uparrow(r)+n_\downarrow(r)$ and the magnetization
$m(r)=n_\uparrow(r)-n_\downarrow(r)$, one should know the equation
of state, $n_\sigma(r)=n_\sigma(\mu(r),h)$ with
$\sigma=\uparrow,\downarrow$. Using the EOS calculated in Section
III, we can determine $\mu_0$ and $h$ from the known total
particle number $N$ and the global polarization
$P=(N_\uparrow-N_\downarrow)/(N_\uparrow+N_\downarrow)$,
\begin{eqnarray}
N=2\pi\int rdr n(r),\ \ \  PN=2\pi\int rdr m(r).\label{number}
\end{eqnarray}

Let us firstly consider a non-interacting system with balanced
populations, $N_\uparrow=N_\downarrow$, which can help us to
define the Fermi energy $\epsilon_{\text F}$ for trapped 2D
system. $\epsilon_{\text F}$ is defined as the chemical potential
$\mu_0$ at the center of the trap for non-interacting gas. The
density profile is also balanced,
$n_{\uparrow}(r)=n_{\downarrow}(r)$, and is given by
\begin{eqnarray}
n_\uparrow(r)=n_\downarrow(r)=c\left(\epsilon_{\text
F}-\frac{1}{2}M\omega^2r^2\right),
\end{eqnarray}
which vanishes at the so called Thomas-Fermi radius
\begin{eqnarray}
R_{\text{T}}=\sqrt{\frac{2\epsilon_{\text
F}}{M\omega^2}}.\label{TF}
\end{eqnarray}
The total density $N$ is then given by the integral
\begin{eqnarray}
N=2\pi \int_0^{R_{\text T}}rdr2c\left(\epsilon_{\text
F}-\frac{1}{2}M\omega^2r^2\right)=\left(\frac{\epsilon_{\text
F}}{\hbar\omega}\right)^2.\label{FermiE}
\end{eqnarray}
We find that the Fermi energy in 2D is $\epsilon_{\text
F}=\sqrt{N}\hbar\omega$, in contrast to the result
$\epsilon_{\text F}=(6N)^{1/3}\hbar\omega$ in 3D (note that we
have recovered $\hbar$ in these expressions).

We then turn to an attractive Fermi gas. For balanced populations,
the ground state is a superfluid state, and the density profile
can be obtained from equation (\ref{EOSSF}),
\begin{eqnarray}
n_\uparrow(r)=n_\downarrow(r)=c\left(\mu_0+\frac{\epsilon_{\text
B}}{2}-\frac{1}{2}M\omega^2r^2\right).
\end{eqnarray}
Comparing with the non-interacting gas, we find no difference
between the normal and the superfluid states. The chemical
potential at the center of the trap reads
\begin{eqnarray}
\mu_0=\sqrt{N}\hbar\omega-\frac{\epsilon_{\text
B}}{2}=\epsilon_{\text F}-\frac{\epsilon_{\text B}}{2}.
\end{eqnarray}
This relation is exactly the same as in the homogeneous
case\cite{2Da}. The order parameter profile $\Delta(r)$ is given
by
\begin{eqnarray}
\Delta(r)=\Delta_0\sqrt{1-r^2/R^2_{\text T}},\ \ \
\Delta_0=\sqrt{2\epsilon_{\text B}\epsilon_{\text F}}.
\end{eqnarray}

For a system with population imbalance, $N_\uparrow\neq
N_\downarrow$, we should have $h\neq0$. By comparing ${\cal
P}_{\text{SF}}$ with ${\cal P}_{\text{N}}$, a first order phase
transition from the phase SF to the phase N occurs for a given
$\mu$ when the Zeeman field $h$ becomes larger than the critical
value $h_c(\mu)$, or equivalently speaking, for a given $h$ the
SF-N phase transition happens when the chemical potential $\mu$
becomes less than the critical value $\mu_c(h)$. In LDA, the phase
behavior as a function of chemical potential $\mu$ is translated
into a spatial cloud profile through $\mu(r)$. The critical phase
boundary $\mu_c$ corresponding to the critical radius $r_c$ is
defined by
\begin{eqnarray}
\mu_c=\mu(r_c)=\mu_0-\frac{1}{2}M\omega^2r_c^2
\end{eqnarray}
at which the states SF and N have the same pressure. Thus, at
fixed $h$, any region of the system which satisfies $\mu(r)>\mu_c$
is in the state SF, while a region which satisfies $\mu(r)<\mu_c$
will be in the state N. Since $\mu(r)$ decreases with increasing
$r$, it is clear that the high density superfluid region will be
confined in the center of the trap, and the low density polarized
state N is expelled to the outside. The shell structure with
radius $r_c$ of the SF-N interface is a striking signature of
phase separation in a trap. The superfluid core will disappear
when the population imbalance $P$ becomes larger than the critical
value $P_c$ which is determined by the equation $r_c=0^+$ or
$\mu_0=\mu_c$.

We should have two types of shell structure corresponding to the
cases $h>h_0$ and $h<h_0$. For the case $h>h_0$, we have
$\mu(r_c)=\mu_c=h^2/\epsilon_{\text B}-\epsilon_{\text B}/4>h$,
which means that there exists a shell of partially polarized
normal gas in the region $r_c<r<r_0$, with $r_0$ given by
$\mu(r_0)=h$. We call it the phase PS-I. Thus we have the
following density profile
\begin{equation}
n(r)=
 \left\{ \begin{array}
{r@{\quad,\quad}l}
 2c\left(\mu_0+\frac{\epsilon_{\text B}}{2}-\frac{1}{2}M\omega^2r^2\right)&
 0<r<r_c \\
 2c\left(\mu_0-\frac{1}{2}M\omega^2r^2\right) & r_c<r<r_0\\
 c\left(\mu_0+h-\frac{1}{2}M\omega^2r^2\right) &
 r_0<r<R
\end{array}
\right.
\end{equation}
and
\begin{equation}
 m(r)=
 \left\{ \begin{array}
{r@{\quad,\quad}l}
 0&
 0<r<r_c \\
 2ch & r_c<r<r_0\\
 c\left(\mu_0+h-\frac{1}{2}M\omega^2r^2\right) &
 r_0<r<R
\end{array}
\right.
\end{equation}
where $R=\sqrt{2(\mu_0+h)/M\omega^2}$ is the edge of the cloud.
After some algebra according to the equation (\ref{number}),
$\mu_0$ is simply given by $\mu_0=\epsilon_{\text
F}-\epsilon_{\text B}/2$ as in the balanced case and $h$ is
solved from the cubic equation
\begin{eqnarray}
2h\left(\frac{h^2}{\epsilon_{\text B}}-\frac{\epsilon_{\text
B}}{4}\right)=P\epsilon_{\text F}^2,
\end{eqnarray}
where $\epsilon_{\text F}=\sqrt{N}\hbar\omega$ is the Fermi energy
defined in (\ref{FermiE}).

From the condition $h>h_0$ which ensures $r_0>r_c$, we have
$P\epsilon_{\text F}^2>2h_0^2$, which leads to the relation
\begin{eqnarray}
P>P_0=\frac{3+2\sqrt{2}}{2}\eta^2.
\end{eqnarray}
The critical polarization $P_c$ is determined by the condition
$\mu_0=\mu_c$. A simple algebra gives
\begin{eqnarray}
P_c=(2-\eta)\sqrt{\eta-\frac{\eta^2}{4}}, \ \ \ \ \ 0<\eta<\eta_1
\end{eqnarray}
with $\eta_1=2-\sqrt{2}\simeq0.586$. Note that both $P_0$ and
$P_c$ reach unity at $\eta=\eta_1$, they are the two boundaries of
the phase PS-I in the $\eta-P$ plane.

For the case $h<h_0$ or $P<P_0$, we have
$\mu(r_c)=\mu_c=(\sqrt{2}h-\epsilon_{\text B})/(2-\sqrt{2})<h$,
which means that the normal gas shell outside the superfluid core
is fully polarized. The density profile reads
\begin{equation}
n(r)=
 \left\{ \begin{array}
{r@{\quad,\quad}l}
 2c\left(\mu_0+\frac{\epsilon_{\text B}}{2}-\frac{1}{2}M\omega^2r^2\right)&
 0<r<r_c \\
 c\left(\mu_0+h-\frac{1}{2}M\omega^2r^2\right) &
 r_c<r<R
\end{array}
\right.
\end{equation}
and
\begin{equation} m(r)=
 \left\{ \begin{array}
{r@{\quad,\quad}l}
 0&
 0<r<r_c \\
 c\left(\mu_0+h-\frac{1}{2}M\omega^2r^2\right) &
 r_c<r<R
\end{array}
\right.
\end{equation}
After the integration in equation (\ref{number}), we still have
$\mu_0=\epsilon_{\text F}-\epsilon_{\text B}/2$ and $h$ is
explicitly given by
\begin{eqnarray}
h=(\sqrt{2}-1)\sqrt{P}\epsilon_{\text F}+\frac{\epsilon_{\text
B}}{2}.
\end{eqnarray}
One can easily check that the condition $h<h_0$ is equivalent to
$P>P_0$, and we have $P_c=1$ for $\eta>\eta_1$.

\begin{figure}[!htb]
\begin{center}
\includegraphics[width=8cm]{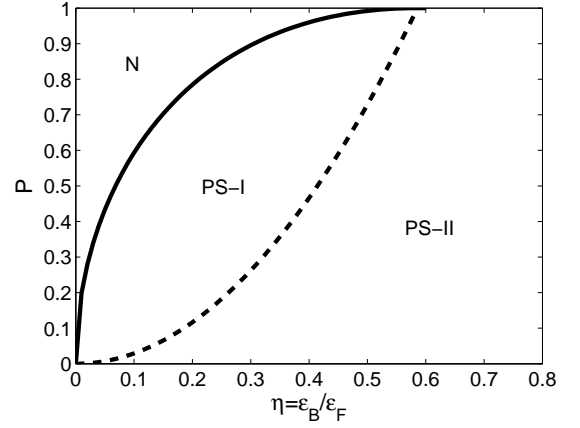}
\caption{Global phase diagram for trapped 2D Fermi gas in the
$\eta-P$ plane. \label{fig4}}
\end{center}
\end{figure}

Fig.\ref{fig4} summarizes the the coupling-imbalance phase diagram
for two-dimensional imbalanced Fermi gas in a harmonic trap. The
critical polarization $P_c=(2-\eta)\sqrt{\eta-\eta^2/4}$ (solid
line) increases from $P_c=0$ at $\eta=0$ to $P_c=1$ at
$\eta=\eta_1\simeq0.586$ and then keeps as a constant $P_c=1$ for
$\eta>\eta_1$. The dashed line, analytically given by
$P=(3+2\sqrt{2})\eta^2/2$, separates the two types of phase
separation, PS-I and PS-II with different shell structure. In the
phase PS-I, the density profile exhibits a
SF-N$_{\text{PP}}$-N$_{\text{FP}}$ shell structure, while in the
phase PS-II, the shell is in the form of SF-N$_{\text{FP}}$.

\begin{figure}[!htb]
\begin{center}
\includegraphics[width=8cm]{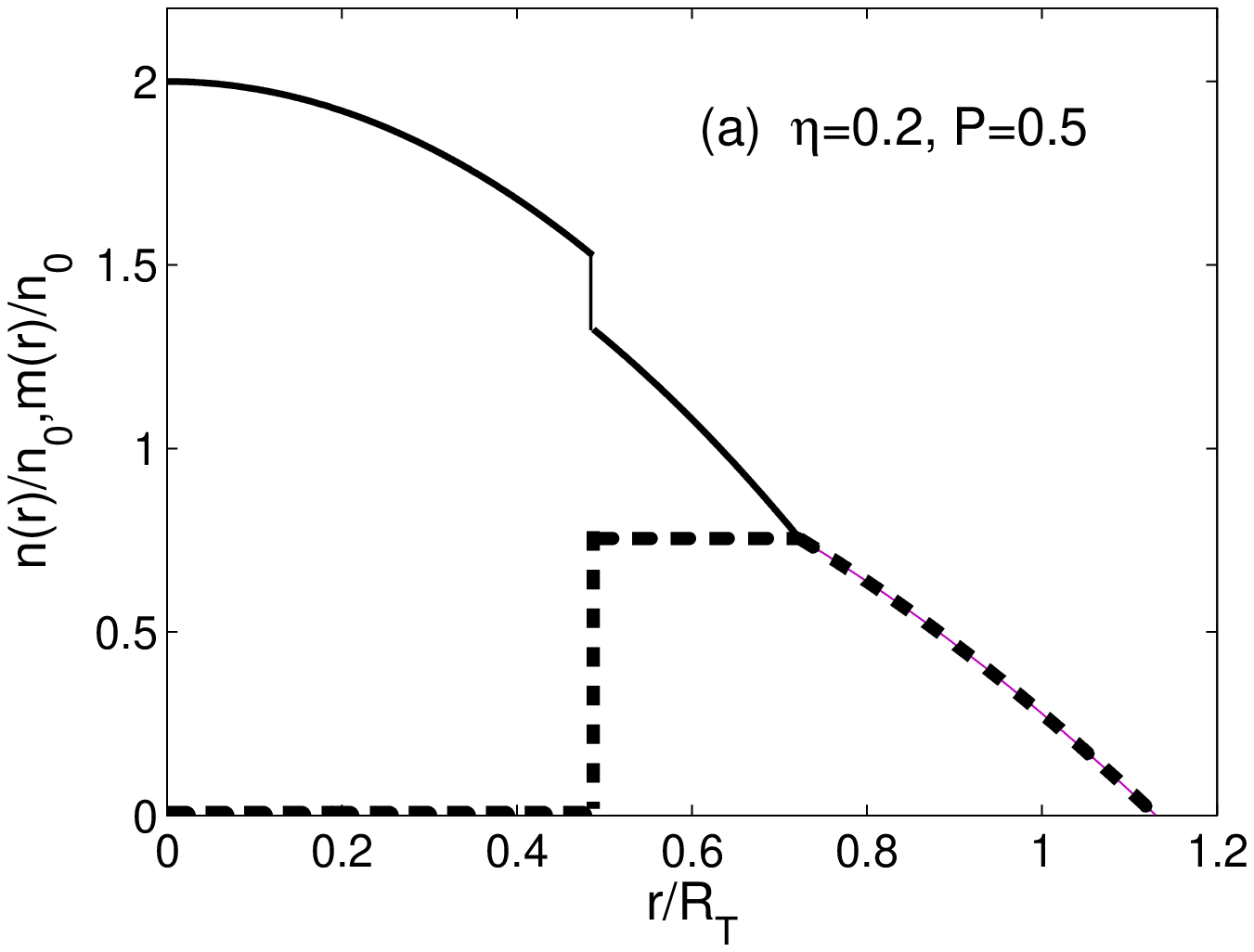}
\includegraphics[width=8cm]{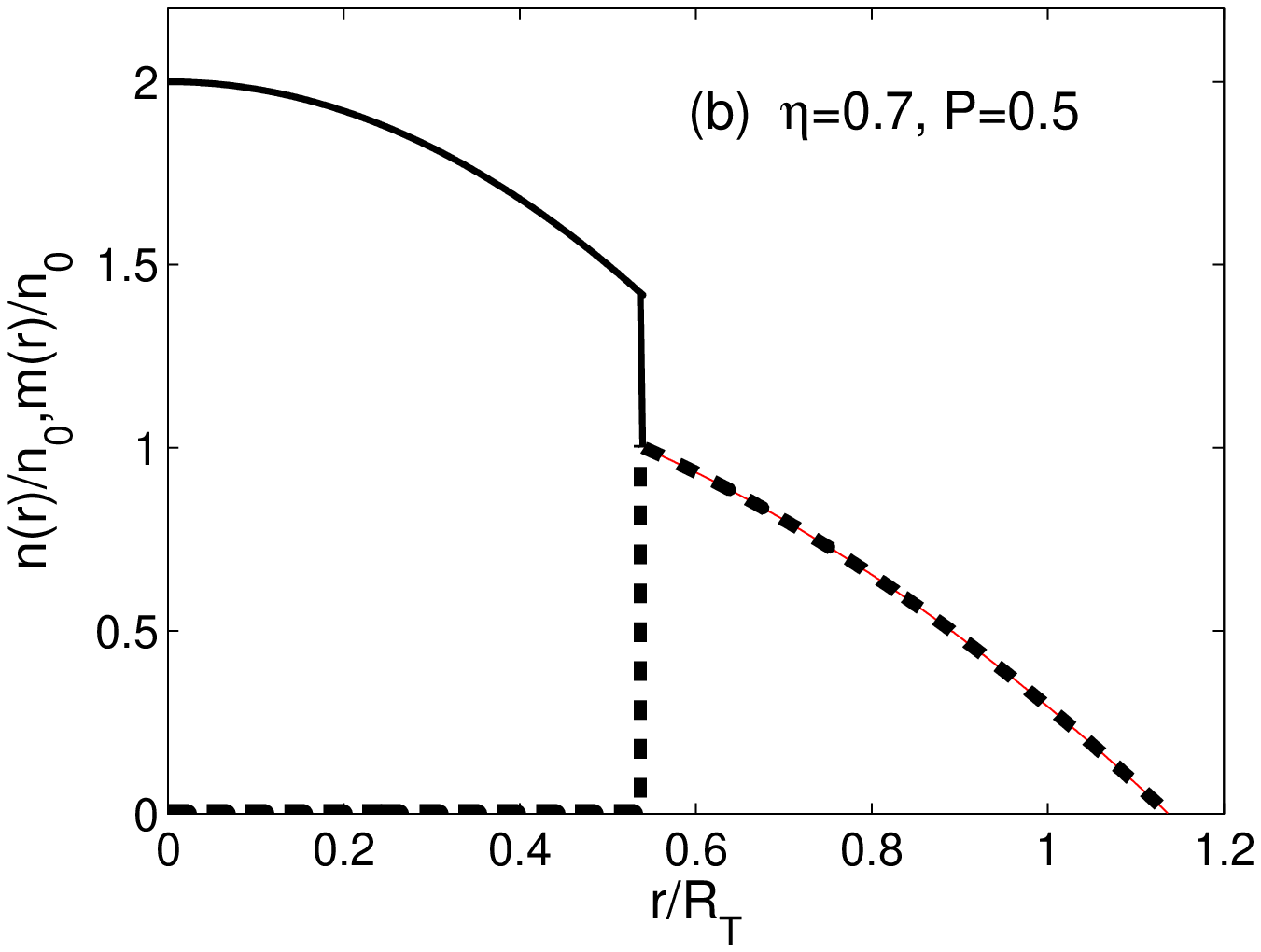}
\caption{The profiles for the total density $n(r) $(solid line)
and magnetization $m(r)$ (dashed line) in two cases, $\eta=0.2$
and $P=0.5$ in the region PS-I and $\eta=0.7$ and $P=0.5$ in the
region PS-II. \label{fig5}}
\end{center}
\end{figure}

The analytical result of the density profile can be summarized as
follows. In the region PS-I, we have
\begin{equation}
\frac{n(r)}{n_0}=
 \left\{ \begin{array}
{r@{\quad,\quad}l}
 2(1-x^2)&
 0<x<x_c \\
 2\left(1-\frac{\eta}{2}-x^2\right) & x_c<x<x_0\\
 1-\frac{\eta}{2}+\delta-x^2 &
 x_0<x<X
\end{array}
\right.
\end{equation}
and
\begin{equation} \frac{m(r)}{n_0}=
 \left\{ \begin{array}
{r@{\quad,\quad}l}
 0&
 0<x<x_c \\
 2\delta & x_c<x<x_0\\
 1-\frac{\eta}{2}+\delta-x^2 &
 x_0<x<X
\end{array}
\right.
\end{equation}
with $n_0=c\epsilon_{\text F}$, $x=r/R_{\text T}$, $R_{\text T}$
being the Thomas-Fermi radius of non-interacting gas defined in
({\ref{TF}), and $\delta=h/\epsilon_{\text F}$ being the real
solution of the cubic equation
$\delta^3-\eta^2\delta/4-P\eta/2=0$,
\begin{eqnarray}
\delta&=&\left(\frac{P\eta}{4}\right)^{1/3}\left[\left(1+\gamma\right)^{1/3}+\left(1-\gamma\right)^{1/3}\right]
\end{eqnarray}
with $\gamma=\sqrt{1-\eta^4/(108P^2)}$. The scaled radii
$x_c=r_c/R_{\text T},x_0=r_0/R_{\text T}$ and $X=R/R_{\text T}$
are given by
\begin{eqnarray}
x_c&=&\sqrt{1-\frac{\eta}{2}-\frac{P}{2\delta}},\nonumber\\
x_0&=&\sqrt{1-\frac{\eta}{2}-\delta},\nonumber\\
X&=&\sqrt{1-\frac{\eta}{2}+\delta}.
\end{eqnarray}
A numerical sample for $\eta=0.2,P=0.5$ is shown in
Fig.\ref{fig5}(a). There is an interesting phenomenon which is
different from that found in 3D: The magnetization profile $m(r)$
exhibits a visible platform structure in the partially polarized
normal shell in the region $r_c<r<r_0$.  For partially polarized
gas, the interaction may be important, like the finding around the
unitary region in 3D~\cite{unitary,MC1,MC2}. However, for the 2D
system, since partially polarized normal shell appears only at
small coupling where the effect of interaction is not important,
our conclusion will not be qualitatively changed. We also observe
a density jump $\Delta n$ at the critical radius $r_c$. In the
region PS-I, $\Delta n$ is independent of the global polarization
and depends only on the coupling strength,
\begin{eqnarray}
\Delta n=\eta n_0=\frac{M}{2\pi}\epsilon_{\text B}.
\end{eqnarray}
Thus the experimental data for $\Delta n$ can be used to extract
the effective two-body binding energy $\epsilon_{\text B}$.

In the region PS-II, the density profile reads
\begin{equation}
\frac{n(r)}{n_0}=
 \left\{ \begin{array}
{r@{\quad,\quad}l}
 2(1-x^2)&
 0<x<x_c \\
 1+(\sqrt{2}-1)\sqrt{P}-x^2 &
 x_c<x<X
\end{array}
\right.
\end{equation}
and
\begin{equation} \frac{m(r)}{n_0}=
 \left\{ \begin{array}
{r@{\quad,\quad}l}
 0&
 0<x<x_c \\
 1+(\sqrt{2}-1)\sqrt{P}-x^2 &
 x_c<x<X
\end{array}
\right.
\end{equation}
The scaled radii $x_c$ and $X$ now takes very simple form
\begin{eqnarray}
x_c&=&\sqrt{1-\sqrt{P}},\nonumber\\
X&=&\sqrt{1+(\sqrt{2}-1)\sqrt{P}}.
\end{eqnarray}
A numerical sample of the density profile for $\eta=0.7,P=0.5$ is
shown in Fig.\ref{fig5}(b). It is very surprising that the density
profile does not depend on the coupling parameter $\eta$, but only
on the global polarization $P$. As a result, the ratio $r_c/R$
exhibits a universal behavior when $\eta>\eta_1=0.586$, as shown
in Fig.\ref{fig6}.
\begin{figure}[!htb]
\begin{center}
\includegraphics[width=8cm]{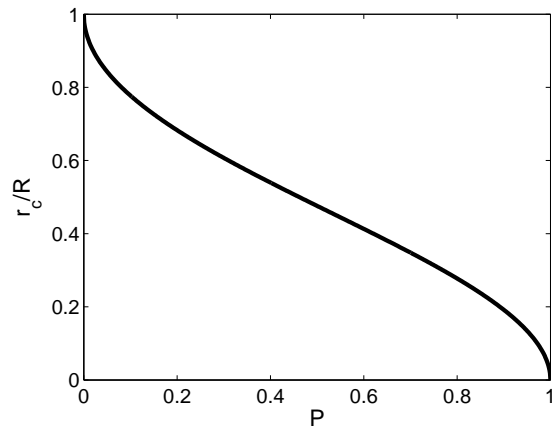}
\caption{The ratio of the superfluid radius to the cloud radius as
a function of the global polarization $P$ for $\eta>\eta_1=0.586$.
\label{fig6}}
\end{center}
\end{figure}

Finally, two comments on our results should be made. The first is
on the BCS-Leggett mean field theory. In this theory, the quantum
fluctuation in the superfluid phase and the interaction in the
partially polarized phase are totally neglected. The effect of
interaction in the partially polarized phase may change the
platform structure in the region $r_c<r<r_0$. However, since the
three-shell structure appears in the weak coupling region, we
expect this effect to be small. At very strong coupling, the
correction in the superfluid phase due to quantum fluctuation
should be important, and the universal behavior in the region
PS-II may be destroyed. Since the BEC region is reached at
$\eta>2$, we expect that our conclusion holds at the BCS side
$\eta_1<\eta<2$. The second comment is on the model we used.
Recently, it is argued that the model we used is not sufficient to
discuss BCS-BEC crossover in quasi-2D Fermi gas due to the
importance of dressed molecules\cite{zhang}. However, from the
study in \cite{zhang}, this effect is important only at strong
coupling (may be for $\eta>2$). Obviously, the comparison of our
prediction with the experimental data can tell us whether the
quantum fluctuation, dressed molecules and other possible effects
are important.

\section {Summary}
In summary, mean field phase structure of polarized Fermi gas in
2D is analytically investigated.  In the normal phase, the
di-fermion bound states at high polarization are easier to survive
than in the balanced Fermi sea. In the BCS-Leggett mean field
theory, the transition from the unpolarized superfluid phase to
the normal phase is always of first order, and there exists no
stable gapless superfluid phase. In the homogeneous system, we
analytically determined the critical Zeeman fields and the
critical population imbalance in the whole coupling parameter
region. We found two critical Zeeman fields in the BCS-BEC
crossover, and proved that the mixed superfluid-normal phase is
the energetically favored ground state. However, from recent
Monte-Carlo simulations~\cite{MC1,MC2}, our mean field results may
be only qualitatively correct in some parameter region, due to the
importance of interactions in the normal phase.

To compare our theoretical results with future experimental data,
we have also calculated analytically the density profile for an
imbalanced 2D Fermi gas confined in a harmonic trap. For balanced
populations, the density profiles for normal and superfluid matter
are the same and can not be used as a signature of superfluidity.
For imbalanced populations, the density profile exhibits a shell
structure, a superfluid core in the center and a normal shell
outside. At small coupling, there exists a partially polarized
normal shell and the density difference shows a platform
structure. For large attraction, however, the normal shell is
fully polarized, and the density profile depends only on the
global population imbalance. Our theoretical prediction can be
examined in the future experiments on 2D ultracold Fermi gases,
which can help us to see whether quantum fluctuations and other
possible effects are important in determining the phase
structure\cite{zhang}.

{\bf Acknowledgments:}\ The work is supported by the NSFC Grants
10575058 and 10735040 and the National Research Program Grant
No.2006CB921404.

\end{document}